\documentclass[12pt]{article}
\usepackage{amsmath,amsthm,amsfonts,amssymb}
\def\qed{{\unskip\nobreak\hfil\penalty50
\hskip2em\hbox{}\nobreak\hfil$\square$
\parfillskip=0pt \finalhyphendemerits=0\par}\medskip}
\def\proof{\trivlist \item[\hskip \labelsep{\it Proof.\ }]}
\def\endproof{\null\hfill\qed\endtrivlist}

\DeclareMathOperator\Aut{Aut}
\DeclareMathOperator\Ad{Ad}

\def\o{{\rm opp}}

\def\a{\alpha}
\def\b{\beta}
\def\d{\delta}
\def\D{\Delta}

\def\epsilon{\varepsilon}
\def\g{\gamma}

\def\f{\varphi}

\def\th{\theta}
\def\o{\omega}
\def\O{\Omega}

\newtheorem{theorem}{Theorem}
\newtheorem{lemma}[theorem]{Lemma}

\newtheorem{corollary}[theorem]{Corollary}

\newtheorem{proposition}[theorem]{Proposition}

\def\2#1{{\cal #1}}

\def\A{{\mathfrak A}}

\def\M{{\mathfrak M}}

\def\H{{\cal H}}

\def\R{{\mathbb R}}
\title{{\bf Graded KMS--Functionals\\
and the Breakdown of Supersymmetry}}

\author{{\sc Detlev Buchholz}\\
Institut f\"ur Theoretische Physik, Universit\"at G\"ottingen,\\
Bunsenstra\ss e 9, D-37073 G\"ottingen, Germany\\
e-mail: buchholz@theorie.physik.uni-goettingen.de\\
\vphantom{X}\\
{\sc Roberto Longo}\footnote{Supported in part by GNAFA and
MURST.} \\
Dipartimento di Matematica,
Universit\`a di Roma ``Tor Vergata''\\
Via della Ricerca Scientifica, I-00133 Roma, Italy\\
e-mail: longo@mat.uniroma2.it}
\date{May 12, 1999}
\begin{document}
\maketitle
\bigskip
\bigskip
\bigskip
\begin{abstract}
\noindent It is shown that the modulus of
any graded or, more generally, twisted
KMS--functional of a C$^*$--dynamical system
is proportional to an ordinary KMS--state and the twist
is weakly inner in the corresponding
GNS--re\-pre\-sen\-ta\-tion. If the
functional is invariant under the adjoint
action of some asymptotically
abelian family of automorphisms,
then the twist is trivial.
As a consequence, such functionals do not
exist for supersymmetric C$^*$--dynamical
systems. This is in contrast with the situation in compact
spaces where super KMS--functionals occur as
super-Gibbs functionals.
\end{abstract}
\newpage
\section{Introduction}
Graded KMS--functionals play a prominent role, both, in physics
and in mathematics. In physics they are used as a tool
in the construction of supersymmetric quantum
field theories in a thermal background \cite{Fu,vH}.
In mathematics they appear in non-commutative geometry, 
notably in the context of Connes'
cyclic cohomology \cite{JaLeOs,Ka} and of the Witten
index \cite{Wi}.

It is the aim of the present article to exhibit in the
general setting of C$^*$--dynamical systems some
elementary properties of graded (or, more generally, twisted)
KMS--functionals which seem to have escaped observation
so far.

The first part of our analysis is complementary
to the work of Stoytchev \cite{St}, who proved that
any normal, faithful and symmetric functional on
a von Neumann algebra is a graded KMS--functional
with respect to the action of some involution and
some canonically associated (modular) automorphism
group. We will show here in the
C$^*$--algebraic setting that the modulus of any twisted KMS functional is
a multiple of an ordinary KMS--state and the twist is weakly inner
in the corresponding GNS--representation.

This structure is familiar from numerous concrete examples
of supersymmetric dynamical systems in compact space.
But, as is shown in the second part of our article, it
disappears if one adds the assumption that the twisted
functional is invariant under the action of some asymptotically
abelian family of automorphisms,
which is typical of infinite systems (thermodynamic limit).
Namely, the twist becomes trivial in this case and the
functional satisfies the ordinary KMS condition. It is
a simple consequence of this result that such functionals
cannot be accommodated in supersymmetric theories.

These results provide further evidence to the effect that thermal
systems can be supersymmetric only in compact space.
For if the spacetime admits a group of symmetries shifting
points spacelike
to infinity, then supersymmetric thermal states can never be
homogeneous with respect to that action.

That supersymmetry is extremely vulnerable to thermal effects
in infinite systems was first pointed out in a model independent
setting in \cite{BuOj}, where it was shown that
supersymmetry is necessarily broken in all spatially homogeneous
KMS states. These results were carried over to a
C$^*$--algebraic setting and
generalized in \cite{Bu}.

The present results show that
graded KMS functionals, which are frequently taken as
building blocks in the construction of supersymmetric
models, would not only break supersymmetry if one
proceeds to the thermodynamic limit \cite{BuOj},
they simply cease to exist.
Some implications of this observation for the study of
infinite dynamical systems are discussed in the conclusions.
\section{Twisted KMS functionals}
Let $\A$ be a unital C*-algebra, $\a$ a one-parameter automorphism
group acting on $\A$, and $\g$ an
automorphism of $\A$. Thus ($\A, \a$) is a
(not necessarily continuous) C$^*$--dynamical system
and $\g$ defines a {\em twist} on $\A$. We are primarily
interested in the case $\g^2={\rm id}$, i.e.\ where $\g$ is a $\mathbb
Z_2$-grading on $\A$, but it will be useful not to assume this
from the outset.

We shall say that a bounded, linear (not necessarily positive)
functional $\f$ on $\A$ is a $\g$-{\it twisted KMS functional}
(or simply a twisted KMS functional)
if, for any given $a,b\in\A$, there
exists a complex function $F_{a,b}\in A(S)$ with
\begin{equation}
        \begin{split}
        F_{a,b}(t)&=\f(\a_t(a)b)\\
        F_{a,b}(t+i)&=\f(\g(b)\a_t(a))\ , \quad t\in \R\ .
        \label{twisted kms}\end{split}
\end{equation}
Here $A(S)$ is the set of bounded continuous functions on the
strip \mbox{$S\equiv\{0\leq\Im z\leq 1\}$}
which are analytic in the interior of $S$.
In physics the width of the strip $S$ has the meaning of
inverse temperature which we have normalized here to $1$ for
convenience.

The same argument as for ordinary ($\g={\rm id}$) KMS states implies
that a $\g$-twisted KMS
functional $\f$  is $\a$-invariant: Since $F_{a,1}(t)=F_{a,1}(t+i)$
the function $F_{a,1}$ extends to a bounded entire function
and therefore is constant.
Setting $a=1$ in (\ref{twisted kms}),
we also see that $\f$ is $\g$-invariant.

Given a $\g$-twisted KMS functional $\f$, we consider its
modulus $\o \equiv |\f|$ which is obtained by extending $\f$
to the second dual of $\A$ and subsequent polar decomposition.
Equivalently, $\o$ may be characterized
\cite[Sec.\ 12.2.9]{Di} as the {\em unique}
positive linear functional on $\A$ which satisfies
$||\o|| = ||\f||$ and
\begin{equation}
|\f (a)|^2 \leq ||\f|| \, \o(a^* a), \quad a \in \A \ . \label{modulus}
\end{equation}
{}From the latter characterization one sees that there holds
$|\f \cdot \b| = \o \cdot \b$ for any automorphism
$\b \in \Aut\A$. Hence if $\b$
preserves $\f$, i.e.\ $\f \cdot \b = \f$, then it also preserves $\o$.

Proceeding to the GNS--representation $\{\pi,\H,\O\}$ of
$\A$ induced by $\o$, one can implement any automorphism
$\b$ which preserves $\f$ by a unitary operator $U_\b$ on $\H$. It is
determined by
\begin{equation}
U_\b \, a \O \equiv \b(a) \O\ , \quad a\in\A\ . \label{implementation}
\end{equation}
We may assume that $\pi$ is $1-1$ (replacing $\A$  by
$\A/\textnormal{ker}\pi$ if necessary), therefore, here and in the
subsequent discussion, we identify $\A$
with its image $\pi(\A)$ under the homomorphism $\pi$
in order to simplify the notation. With this convention,
$\b$ extends to an automorphism $\widetilde\b$ of the weak closure
$\M = \A^{\prime \prime}$ given by
$\widetilde\b  = \Ad U_\b $.

As is well known (and can be seen from (\ref{modulus}))
the functional $\f$ can be represented in
the form (polar decomposition)
\begin{equation}
\f  =  (u \, \cdot \, \O, \O). \label{polar}
\end{equation}
Here $u \in \M$  is a partial isometry which is uniquely fixed
by the condition that $u u^*$ is the support projection of
$\o$, i.e.\ the smallest projection $p \in \M$ for which
$p \O = \O$.

We consider in the following the canonical extensions
of $\o$, $\f$ to $\M$, which are given by
$\widetilde\o\equiv (\, \cdot\, \O,\O)$  and $\widetilde\f\equiv
(u\, \cdot \, \O,\O)$, respectively. The following result can be
established by standard arguments, cf. \cite{Ta}.

\begin{lemma}\label{ext} $\tilde\f$ is a $\tilde\g$-twisted KMS
functional of $(\M, \tilde\a)$.
\end{lemma}
\proof
If $x,y\in\M$, there exist by Kaplansky's density theorem
bounded nets $a_i\in\A$ and $b_j\in\A$ such that $a_i \to x$
and $b_j \to y$ $^*$-strongly. The corresponding family of functions
$\{t \mapsto \f (\a_t(a_i) b_j)\}_{i,j}$  converges
to $t \mapsto \widetilde{\f} (\widetilde{\a}_t(x)y)$ uniformly on $\R$,
as can be seen from the estimate
\begin{equation}
| {\f} ({\a}_t(a_i) b_j) -
\widetilde{\f} (\widetilde{\a}_t(x)y) |
\leq ||(a_i - x)^* u^* \O|| \, || b_j \O|| + ||(b_j - y) \O || \,
||x^* u^* \O || \, .
\end{equation}
Here the $\widetilde{\a}$--invariance of
$\widetilde{\f}\, , \widetilde{\o}$ and the
polar decomposition of $\widetilde{\f}$
have been used. In a similar manner one sees that
the family $\{t \mapsto \f (\g(b_j) \a_t(a_i))\}_{i,j}$ converges
to $t \mapsto \widetilde{\f} (\widetilde{\g}(y) \widetilde{\a}_t(x))$
uniformly on $\R$. This implies, since the maximum modulus principle
holds on $A(S)$ according to the Three--Line--Theorem, that
the family $\{F_{a_i,b_j}\}_{i,j}$ converges uniformly on the
strip $S$ to some function $\widetilde{F}_{x,y}$
which also belongs to $A(S)$. Moreover,
$\widetilde{F}_{x,y} (t) = \widetilde{\f} (\widetilde{\a}_t(x)y)$
and $\widetilde{F}_{x,y} (t + i) = \widetilde{\f} (\widetilde{\g}(y)
\widetilde{\a}_t(x))$.
\endproof
\begin{lemma}\label{triv} Let $\o$ be a $\g$-twisted KMS functional
of $(\A\, , \a)$.
If $\o$ is positive, then $\tilde\g ={\rm id}$ and $\widetilde{\o}$
is an ordinary positive KMS functional of $(\M \, , \widetilde{\a})$.
\label{twist}\end{lemma}
\proof
As $\o$ is positive it coincides with its modulus and the
associated partial isometry $u \in \M$ satisfies $u^* \O = \O$.
We shall show next that $\O$ is separating for $\M$.
Let $U(t)=e^{iHt}$ be the one--parameter unitary group implementing
$\widetilde{\a}_t$, cf.\ relation (\ref{implementation}),
and let $V$ be the unitary operator implementing $\widetilde{\g}$.
According to Lemma \ref{ext}, the functional $\widetilde{\o}$
is a $\widetilde\g$--twisted positive KMS functional for
$(\M\, , \widetilde{\a})$.
Now if $x \in \M$ is such that $x\O=0$, then also
$\widetilde{\a}_t(x)\O=0$, hence, making use of formula (\ref{twisted kms}),
we have for all $y\in\M$
\begin{multline}
x\O=0 \Rightarrow \
\widetilde{\o}(\widetilde{\g}(y) \widetilde{\a}_t(x)) =
(\widetilde{\g}(y)\widetilde{\a}_t(x) \O , \O) = 0 \\
\Rightarrow \ \widetilde{\o}(xy) = 0 \Rightarrow \ (y\O,x^*\O)= 0 
\Rightarrow \ x^*\O=0 \, .
\end{multline}
Thus
$x\O=0\Rightarrow zx\O=0 \, \Rightarrow \, x^*z^*\O=0$ for all $z\in\M$,
hence $x=0$. As an immediate consequence of this observation we have
$u^*=1$. It remains to show that $\widetilde{\g} = \mbox{id}$.

Let $\D^{it}$ be the one--parameter modular group associated with
$\M$ and $\O$. As $\O$ is $U(s)$--invariant and
$\widetilde{\a}_s(\M)=\M$, the unitary groups
$\D^{it}$ and $U(s)$ commute. Thus there exists a dense subalgebra
$\M_0\subset\M$ such that $\M_0\O$ is a core both for
$\D$ and $e^{H}$.
By the KMS property for the modular group we have
\begin{equation}
(x^*\O,y^*\O)=( yx^*\O,\O)=(x^*\D y\O,\O)=(\D y\O,x\O)\ ,\quad
x,y\in\M_0\ .
\end{equation}
On the other hand, by the $\g$--twisted KMS condition, we have
\begin{multline}
(x^*\O, y^*\O) =( yx^*\O,\O) = (\widetilde{\g}(x^*)e^{H}y\O,\O)=
(e^{H}y\O,\widetilde{\g}(x)\O) \\
=  (e^{H}y\O,Vx\O)=(V^*e^{H}y\O,x\O)\ ,\quad  x,y\in\M_0\ .
\end{multline}
Hence $\D= V^*e^{H}$ and, by the uniqueness of the
polar decomposition for closed linear operators,
$\D=e^{H}$ and $V=1$, i.e.\ $\widetilde{\g}={\rm id}$.
\endproof
\begin{lemma} Let $\f$ be a $\g$--twisted KMS functional
of $(\A, \a)$ and let $\f =  (u \, \cdot \, \O, \O)$
be its polar decomposition. Then $u$ is unitary
and $\O$ is separating for $\M$.
\label{separating}
\end{lemma}
\proof
We begin by noting that
$u$ is left fixed both by $\widetilde{\g}$ and
$\widetilde{\a}$ because of the invariance of $\widetilde{\f}$
and $\widetilde{\o}$ under the adjoint action of
these automorphisms and the uniqueness of the polar
decomposition. Now, for $\widetilde{\a}$--analytic
elements $x\in\M$
the twisted KMS condition for $\widetilde{\f}$ can be expressed as
\begin{equation}
\widetilde{\o}(u\widetilde{\a}_i(x)y)=
\widetilde{\o}(u \widetilde{\g}(y) x)\ .
\end{equation}
Replacing $x$ with $u^*x$ and using the
$\widetilde{\a}$--invariance of $u$ as well as the fact that
$u u^*$ is the support projection of
$\widetilde{\o}$, we can proceed to
\begin{equation}\label{om-kms}
\widetilde{\o}(\widetilde{\a}_i(x)y)=
\widetilde{\o}(u\widetilde{\g}(y)u^* x)\ .
\end{equation}
Setting $x=u$ and $y=u^*$ in this formula we get, taking into account
that  $\widetilde{\g}(u^*) = u^*$,
\begin{equation}
\widetilde{\o}(1) =
\widetilde{\o}(u u^*) =
\widetilde{\o}(u u^* u^* u)=
\widetilde{\o}(u^* u)\ ,
\end{equation}
hence $u^* u \O = u u^* \O = \O$ by the limit case of the Schwartz
inequality.

We shall show now that $\O$ is separating for $\M$, and this will follow
as above by showing that $x\O=0$ for $x \in \M$ implies $x^*\O=0$.
So let $x \in \M$ satisfy $x\O=0$. Then
$\widetilde{\f}(y\widetilde{\a}_t(x))=
(uy\widetilde{\a}_t(x)\O, \O)=0$ for all $y\in\M$.
Hence by equation (\ref{twisted kms}) we have
$(u \widetilde{\g}(x)y\O,\O) = \widetilde{\f}(\widetilde{\g}(x) y)
= 0$
and this implies $\widetilde{\g}(x^*)u^*\O=0$ or
equivalently, since $\widetilde{\g}(u^*) = u^*$, that $x^*u^*\O=0$.
As $uu^*$ is the support
projection of $\widetilde{\o}$ we conclude that
$x^*u^*= x^*u^* u u^* = 0$ and consequently
$x^* u^* u = 0$. But $u^* u \O=\O$, so it follows that
$x^*\O=0$, i.e.\ $\O$ is separating. Hence $u^* u = u u^* = 1$.
\endproof
We mention as an aside that the above result allows one to disintegrate
a twisted KMS functional into factorial twisted KMS functionals.

\begin{proposition} \label{phi-structure}
Let $\f$ be a $\g$--twisted KMS functional
of $(\A, \a)$ and let $\f =  (u \, \cdot \, \O, \O)$
be its polar decomposition.
Then $\widetilde\g$ is inner on $\M$, indeed $\widetilde\g=\Ad u^*$, and
$\widetilde{\o}$ is an ordinary positive KMS functional for
$(\M, \widetilde{\a})$.
\end{proposition}
\proof
Since $u$ is unitary we have
$\widetilde{\th} \equiv \Ad u \cdot \widetilde{\g} \in \Aut \M$,
so formula (\ref{om-kms}) shows
that $\widetilde{\o}$ is a $\widetilde{\th}$--twisted
KMS functional of $(\M, \widetilde{\a})$. (We recall that this
formula amounts to relation (\ref{twisted kms})
for $\widetilde{\a}$-entire elements $x \in \M$; that it entails relation
(\ref{twisted kms}) for all elements of $\M$ can be shown by
similar arguments as in the proof of Lemma \ref{ext}.)
As $\widetilde{\o}$ is positive, we conclude from Lemma \ref{triv} that
$\widetilde{\th} = {\rm id}$. Hence $\widetilde{\o}$ is an ordinary
positive KMS functional and $\widetilde{\g} = \Ad u^*$.
\endproof
\begin{corollary} \label{commute}
If $\b\in\Aut\A$ preserves the $\g$--twisted KMS functional
$\f$ of $(\A,\a)$, then $\widetilde\b$ commutes with
$\widetilde\a$ and $\widetilde\g$.
\end{corollary}
\proof
Since $\widetilde\b(u)=u$ by the uniqueness of the polar decomposition
of $\f$, the commutativity of $\widetilde\b$
and $\widetilde\g$ follows from the preceding result.
As $\widetilde{\a}$ is the modular group associated with
$(\M,\O)$, it is also clear that
$\widetilde\b$ commutes with $\widetilde{\a}$.
\endproof
\section{Twisted asymptotic abelianess}
We shall specialize now to the class of C$^*$--dynamical systems
$(\A,\a)$ for which there exists a family of automorphisms
acting on $\A$ in a (twisted)
asymptotically abelian manner. This situation
prevails in physics if one deals with infinite systems
(thermodynamic limit). As a matter of fact, the dynamics $\a$
itself is asymptotically abelian in generic cases.

Given a $\g$-twisted KMS functional $\f$ of $(\A,\a)$,
we shall say that a sequence of automorphisms $\b_n\in\Aut\A$
is {\it $\f$-asymptotically abelian} if $\f\cdot\b_n=\f$ and
\begin{equation}
\lim_n \, \f(c[a,\b_n(b)]) = 0  \quad \mbox{for all} \ a,b,c\in\A\ .
\label{abelianess}
\end{equation}
Here the twisted commutator is defined by $[a,b]\equiv ab- \g(b) a$.
\begin{lemma}
Let $\f$ be a $\g$-twisted KMS functional of $(\A,\a)$. If there exists
a $\f$--asymp\-to\-ti\-cally abelian sequence $\b_n\in\Aut\A$,
then $[x,\widetilde\b_n(y)]\to 0$ weakly for all
$x,y\in\M$.
\end{lemma}
\proof Since $([a,\b_n(b)]\O,c^*u^*\O)=\f(c[a,\b_n(b)])$ and the
set of vectors $c^* u^* \O, c \in \A$, is dense in $\H$ (recall that
$u$ is a unitary in $\M = \A^{\prime \prime}$), it follows from condition
(\ref{abelianess}) that the (bounded) sequence
$[a,\b_n(b)] \O$ converges weakly to $0$. Thus
$[a,\b_n(b)] \rightarrow 0$ weakly because
all weak limit points of  $[a,\b_n(b)]$ are elements of $\M$ and
$\O$ is separating for $\M$ by Lemma \ref{separating}.

As the unit ball $\A_1$ of $\A$ is $^*$-strongly dense in the unit ball
$\M_1$ of $\M$ by Kaplansky's density
theorem, given $x,y\in\M_1$ and $\varepsilon > 0$, there exist
$a,b\in\A_1$ with
\begin{equation}
||x\O-  a\O||, \ ||x^*\O-  a^*\O||, \ ||y\O-  b\O||, \ ||y^*\O-
b^*\O||<\varepsilon\ .
\end{equation}
Now  for any fixed $z'\in\M'_1$ we have
\begin{multline}\label{est}
|([x,\widetilde\b_n(y)]\O,z'\O) - ([a,\widetilde{\b}_n(b)]\O,z'\O)|\\
\leq |\big(( x\widetilde\b_n(y) - a\widetilde\b_n(b)) \O,z'\O\big)|
+ |\big((\widetilde{\g}(\widetilde\b_n(y))x -
\widetilde{\g}(\widetilde\b_n(b))a)\O,z'\O\big)|\ .
\end{multline}
The first term in the right hand side of the above inequality can be
estimated by
\begin{multline}
|\big(( x\widetilde\b_n(y) - a\widetilde\b_n(b)) \O,z'\O\big)|
\leq |\big((x-a)\widetilde\b_n(y)\O,z'\O)|+
|(a\widetilde\b_n(y-b)\O,z'\O\big)|\\
\leq |(\widetilde\b_n(y) \O,z'(x-a)^*\O)|+
|(\widetilde\b_n(y-b)\O,z'a^*\O)|\leq 2\|\O\|^2
\varepsilon\ ,
\end{multline}
where, in the last step, we used the fact that
$\widetilde\b_n$ preserves $\o$.
Similarly, the second term on the right hand side of (\ref{est}) is
bounded by $2\|\O\|^2\varepsilon$.

Since $\varepsilon$ is arbitrary, this entails
$([x,\widetilde\b_n(y)]\O,z'\O)\rightarrow 0$, hence,
as the set of vectors $\M^{\prime}\, \O$ is dense in
$\H$ and $\O$ is separating for $\M$, we conclude that
$[x,\widetilde\b_n(y)] \rightarrow 0$ weakly
for all $x,y\in\M$.
\endproof

\begin{proposition}\label{aa}
Let $\f$ be a $\g$-twisted KMS functional of $(\A,\a)$. If there exists
a $\f$-asym\-pto\-ti\-cally abelian sequence $\b_n\in\Aut\A$, then
$\widetilde\g={\rm id}$ and
$\widetilde{\f}$ is an ordinary KMS functional of $(\M,\widetilde{\a})$.
\end{proposition}
\proof
As noticed above, there holds
$\widetilde{\b}_n(u) = \widetilde{\g}(u)=u$ by
the uniqueness of the polar decomposition of $\f$.
Thus, by the preceding lemma,
\begin{equation}
xu - ux =
x \widetilde{\b}_n(u) - \widetilde{\b}_n(u) x =
x \widetilde{\b}_n(u) - \widetilde{\g}\big( \widetilde{\b}_n(u) \big)x
=[x, \widetilde{\b}_n(u)]\rightarrow 0
\end{equation}
weakly for all $x\in\M$.
So $u$ belongs to the center of $\M$ and, as $\widetilde{\g}=\Ad u^*$
by Proposition \ref{phi-structure}, $\widetilde{\g}$ is trivial.
\endproof
Let us now assume that $\widetilde{\g}^2={\rm id}$.
Note that this is always the case
if the $\g$--twisted  KMS functional $\f$ is selfadjoint,
i.e.\ $\f(a^*)=\overline{\f(a)}$, $a \in \A$. For then
\begin{equation}
\widetilde{\o}(ua) = \f(a)=\overline{\f(a^*)}=
\overline{\widetilde{\o}(ua^*) }
=\widetilde{\o}(au^*)=\widetilde{\o}(u^*a) , \quad a\in\A \ ,
\end{equation}
where in the last equality we made use of the fact that
$\widetilde{\o}$ is a KMS functional and
$\widetilde{\a}_t(u^*)=u^*$. Thus $u=u^*$ since
$\widetilde{\o}$ is faithful, hence
$\widetilde{\g}^2=\Ad {u^*}{}^2 = {\rm id}$. A
$\g$--twisted KMS functional is called {\it graded} KMS functional
if $\widetilde{\g}$ has the latter property.

Given a graded KMS--functional $\f$ of $(\A,\a)$, let
$\M_{\rm an}\subset \M$ be the weakly dense algebra of
$\widetilde{\a}$--analytic elements and let
\begin{equation}
\widetilde{\d}_0\equiv
-i\frac{\mbox{\footnotesize $\rm d$}}{\mbox{\footnotesize ${\rm d}t$}}
\widetilde{\a}_t|_{t=0}
\end{equation}
be the generator of $\widetilde{\a}$, which has $\M_{\rm an}$ as a core.
The dynamical system $(\A,\a)$ is
said to be {\it $\f$--supersymmetric} if there exists a closable
odd derivation
$\widetilde{\d}:\M_{\rm an}\to \M_{\rm an}$ such that
\begin{equation} \label{susy}
\widetilde{\d}^{\,2}(a) = \widetilde{\d}_0(a) \ , \quad a\in\M_{\rm an} \ .
\end{equation}
Here by an odd derivation we mean a graded derivation, i.e.\
\begin{equation}
\widetilde{\d}(ab)=\widetilde{\d}(a)b+
\widetilde{\g}(a)\widetilde{\d}(b)\ ,\quad a,b\in\M_{\rm an} \ ,
\end{equation}
which is odd, i.e.\
$\widetilde{\d}\cdot\widetilde{\g}=-\widetilde{\g}\cdot\widetilde{\d}$
on $\M_{\rm an}$.
Note that the condition of $\f$--supersymmetry
for a C$^*$--dynamical system is somewhat weaker
than the condition of supersymmetry
since the existence of supersymmetry
transformations is only required in the representation induced by
$\f$.

\begin{corollary} Let $\f$ be a graded KMS functional
of the $\f$--supersymmetric C$^*$--dy\-nam\-ical system $(\A,\a)$.
If there exists a $\f$--asymptotically
abelian sequence $\b_n\in\Aut\A$, then
$\widetilde{\d}=0$, $\widetilde{\a} = \mbox{\rm id}$,
and $\widetilde{\f}$ is a trace.
\end{corollary}
\proof
By Proposition \ref{aa}, $\widetilde{\f}$ is an ordinary KMS functional
and the grading $\widetilde{\g}$ is trivial. Since
$\widetilde{\d}$ is odd, it follows that
$\widetilde{\d} = - \widetilde{\d}= 0$ on $\M_{\rm an}$,
hence $\widetilde{\d}_0 = 0$ because of equation (\ref{susy}) and the fact
that $\M_{\rm an}$ is a core for $\widetilde{\d}_0$.
Thus $\widetilde{\a}=\mbox{\rm id}$, hence
by  the KMS property we conclude that
$\widetilde{\f}$ is a trace on $\M$.
\endproof
\section{Conclusions}
In the preceding analysis we have determined the
form of graded KMS functionals of dynamical systems
in a quite general setting and we want to discuss now
some implications of our results for the study of
infinite systems.

In order to fix ideas, let us assume that we
are dealing with a family of graded dynamical systems
($\A_\Lambda, \a_\Lambda , \g_\Lambda$)
which are assigned to compact subsets
$\Lambda \subset \R^n$ and which
determine in the limit $\Lambda \nearrow \R^n$ some
dynamical system ($\A, \a, \g$)
on which spatial translations act in an
asymptotically abelian manner. This situation
is familiar from quantum field theory,
where one frequently constructs first the theory
in a finite volume (box) $\Lambda$ and then proceeds to
the thermodynamic limit.

For the finite volume theory one expects that
the graded KMS functionals
generically give rise to type I representations. They
can then be presented in the familiar Gibbs form,
\begin{equation} \label{Gibbs}
\f_\Lambda (\,\cdot\,) = {\rm Tr} \,
e^{\, - \beta H_\Lambda} \, V_\Lambda \ \cdot \ \ ,
\end{equation}
where $e^{\, - \beta H_\Lambda}$ is the density matrix of the
Gibbs ensemble at inverse temperature $\beta$ and $V_\Lambda$ the
unitary which implements the grading and commutes with the
box--Hamiltonian $H_\Lambda$. If $\Lambda \nearrow \R^n$,
the representation (\ref{Gibbs}) is no longer meaningful, however, since
the limit of $e^{\, - \beta H_\Lambda}$ is in general
not a trace class operator. So the question arises
in which sense a thermodynamic limit of
graded Gibbs functionals can be defined.

{}From the point of view of physics it might seem natural to normalize
the functionals, i.e.\ to  proceed from
$\f_\Lambda$ to $||\f_\Lambda||^{-1} \, \f_\Lambda$.
The normalized functionals
could then be interpreted as weighted differences of
bosonic and fermionic ensembles \cite{vH}
and the existence of (weak--*) limits would
follow from standard compactness arguments. That
this idea does not work in general can be seen
if one thinks of the situation where one has
a unique KMS state for given $\beta$, for example at high
temperatures $\beta^{-1}$.
Then the limit functionals are invariant under the spatial
translations. But this is in conflict
with Proposition 7 according to which the
grading would have to be trivial, unless the limit
is zero. So this approach does not seem viable.

The other obvious idea is to normalize the functionals by fixing the
value of $\f_\Lambda (1)$ (the index), assuming that it
is different from $0$. But then the norms $||\f_\Lambda||$
cannot stay bounded;
for otherwise one would come, as above, to the conclusion that the
limit points of the functionals $\f_\Lambda$ are zero in general,
in conflict with their normalization. Hence in
this approach one has to deal with unbounded sequences of
functionals and it is not clear from the outset how a reasonable
limit can be defined in these cases. To the best of our knowledge,
this problem has not been solved so far in any non--trivial example.

{}From a mathematical point of view the second approach
through unbounded functionals seems nevertheless
attractive. For it provides a natural generalization
of the concept of graded KMS functionals to the
class of asymptotically abelian C$^*$--dynamical systems.
The unboundedness would not be an obstacle
to the definition and analysis of entire cyclic cocyles
\cite{Co} for such systems, should the functionals be continuous with respect
to some auxiliary Banach algebra norm. It is less clear, however, how such
unbounded functionals can be interpreted in physics.

\bigskip

\noindent {\bf \Large Acknowledgements}\\[2mm]
\noindent The second named author wishes to thank Klaus Fredenhagen
for conversations. Both authors
are grateful to the Erwin Schr\"odinger Institut in Vienna
for hospitality and financial support, which made this collaboration
possible.

\bigskip 

\noindent {\bf \Large Note added}\\[2mm]
\noindent The assumptions of Corollary 8 are unnecessarily strong.  In fact,
they would exclude many examples of interest \cite{KiNa}. We thank 
Ola Bratteli for pointing out this reference to us.

The statement holds, however, in more generality if one replaces in the 
definition of $\varphi$--supersymmetric dynamical systems the 
algebra ${\M}_{\mbox{\scriptsize an}}$ by any weakly dense 
$\widetilde{\gamma}$--in\-variant subalgebra  
$\M_0 \subset \M$ which is contained 
in the intersection of the domains of $\widetilde{\delta}_0$ and 
$\widetilde{\delta}^2$.
The existence of such a subalgebra is clearly a necessary prerequisite
for the definition of supersymmetry. 

The proof that $\widetilde{\delta}_0=0$ remains true in 
this more general situation
since $\widetilde{\delta}_0$ is a weakly closed operator. If such an 
operator vanishes on some dense domain, it vanishes identically.
By the original argument, $\widetilde{\delta}_0 \upharpoonright \M_0 = 0$, 
so one arrives at the statement of the corollary 
also under these weak conditions.

\end{document}